%% file: prr_training_robust_quantum_models.tex
\documentclass[%
 reprint,
groupedaddress,
 amsmath,amssymb,
 aps,
pra,
]{revtex4-2}

\usepackage{graphicx}
\usepackage{dcolumn}
\usepackage{bm}
\usepackage{comment}
\usepackage{amsthm}
\usepackage{mathtools}
\usepackage{color}

\newcommand{\JB}[1]{{\color{black}#1}}

\usepackage{braket}
\usepackage{subcaption}

\input{commands_environments}

\begin{document}

\preprint{APS/123-QED}

\title{Training robust and generalizable quantum models}

\author{Julian Berberich$^1$, Daniel Fink$^2$, Daniel Pranji{\'c}$^3$, Christian Tutschku$^3$, and Christian Holm$^2$}
 \affiliation{$^1$University of Stuttgart, Institute for Systems Theory and Automatic Control, 70569 Stuttgart, Germany}
 \email{julian.berberich@ist.uni-stuttgart.de}
 \affiliation{
 $^2$University of Stuttgart, Institute for Computational Physics, 70569 Stuttgart, Germany}
 \affiliation{
 $^3$Fraunhofer IAO, Fraunhofer Institute for Industrial Engineering, 
 70569 Stuttgart, Germany
 }

\date{\today}

\begin{abstract}
    Adversarial robustness and generalization are both crucial properties of reliable  machine learning models.
    In this paper, we study these properties in the context of quantum machine learning based on Lipschitz bounds.
    We derive parameter-dependent Lipschitz bounds for quantum models with trainable encoding, showing that the norm of the data encoding has a crucial impact on the robustness against data perturbations.
    Further, we derive a bound on the generalization error which explicitly involves the parameters of the data encoding.
    Based on these theoretical results, we propose a practical strategy for training robust and generalizable quantum models by regularizing the Lipschitz bound in the cost.
    Moreover, we show that, for fixed and non-trainable encodings, as those frequently employed in quantum machine learning, the Lipschitz bound cannot be influenced by tuning the parameters.
    Thus, trainable encodings are crucial for systematically adapting robustness and generalization during training.
    The practical implications of our theoretical findings are illustrated with numerical results.
\end{abstract}

\maketitle


\section{Introduction}\label{sec:introduction}

Robustness of machine learning (ML) models is an increasingly important property, especially when operating on real-world data subject to perturbations.
In practice, there are various possible sources of perturbations such as noisy data acquisition or adversarial attacks.
The latter are tiny but carefully chosen manipulations of the data, and they can lead to dramatic misclassification in neural networks~\cite{goodfellow2014explaining,szegedy2014intriguing}.
As a result, much research has been devoted to better understanding and improving adversarial robustness~\cite{wong2018provable,tsuzuku2018lipschitz,madry2019towards}. 
It is well-known that robustness is closely connected to generalization~\cite{goodfellow2014explaining,szegedy2014intriguing,krogh1991simple,xu2012robustness,papernot2016distillation}, i.e., the ability of a model to extrapolate beyond the training data.
Intuitively, if a model is robust then small input changes only cause small output changes, thus counteracting the risk of overfitting.

A Lipschitz bound of a model $f$ is any $L>0$ satisfying 
\begin{align}\label{eq:def_lipschitz}
    \lVert f(x_1) -f(x_2)\rVert 
    \leq L\lVert x_1 - x_2\rVert
\end{align}
for all $x_1,x_2 \in \mathcal{D} \subseteq \mathbb{R}^d$, \JB{where $d$ is the data dimension}.
By definition, Lipschitz bounds quantify the worst-case output change that can be caused by data perturbations and, thus, they provide a useful measure of adversarial robustness.
Therefore, they are a well-established tool for characterizing robustness and generalization properties of ML models~\cite{szegedy2014intriguing,krogh1991simple,luxburg2004distance,bartlett2017spectrally,neyshabur2017exploring,sokolic2017robust,weng2018evaluating,ruan2018reachability,wei2019data}.
Lipschitz bounds cannot only be used to better understand these two properties, but they also allow one to improve them by regularizing the Lipschitz bound during training~\cite{szegedy2014intriguing,krogh1991simple,hein2017formal,gouk2021regularisation,pauli2022training}.

\begin{figure*}
    \begin{center}
        \includegraphics[scale=0.658]{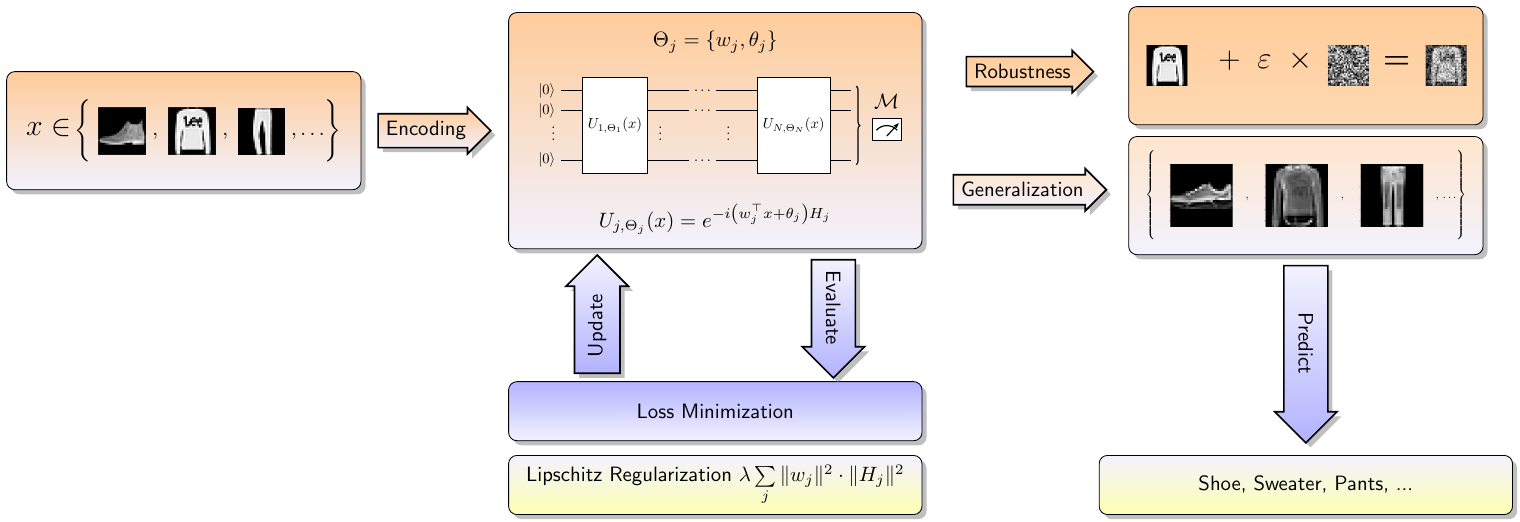}	
    \end{center}
    \caption{Schematic illustration of the quantum model and training setup considered in this work for an exemplary Fashion MNIST data set~\cite{xiao2017fashionmnist}.
    The data, $x$, enter the quantum circuit via a trainable encoding, i.e., they are encoded into unitary operators $U_{j,\Theta_j}(x)$ via an affine function $w_j^\top x+\theta_j$ with trainable parameters $w_j$, $\theta_j$.
    During training, we minimize a cost function consisting of the empirical loss as well as an additional regularization term penalizing the norms of the parameters $w_j$.
    This regularization reduces the Lipschitz bound of the quantum model w.r.t.\ data perturbations \JB{and, thereby, encourages improved robustness and generalization properties.}}
    \label{fig:overview}
\end{figure*}

In this paper, we study the interplay of robustness and generalization in quantum machine learning (QML).
Variational quantum circuits are a well-studied class of quantum models~\cite{schuld2021machine,benedetti2019parameterized,havlicek2019supervised,schuld2019quantum} and they promise benefits over classical ML in various aspects including trainability, expressivity, and generalization performance~\cite{abbas2021power,jerbi2023quantum}.
Data re-uploading circuits generalize the classical variational circuits by concatenating a data encoding and a parametrized quantum circuit not only once but repeatedly, thus iterating between data- and parameter-dependent gates~\cite{perez2020data}.
This alternation provides substantial improvements on expressivity, leading to a universal quantum classifier even in the single-qubit case~\cite{perez2020data,schuld2021effect}.

Just as in the classical case, robustness is crucial for quantum models.
First, if QML is to provide benefits over classical ML, it is necessary to implement QML circuits which are robust w.r.t.\ quantum errors occurring due to imperfect hardware in the noisy intermediate-scale quantum (NISQ) era~\cite{preskill2018quantum}.
Questions of robustness of quantum models against such hardware errors have been studied, e.g., in~\cite{larose2020robust,cincio2021machine}.
Lipschitz bounds can be used to study robustness of quantum algorithms against certain types of hardware errors, e.g., coherent control errors~\cite{berberich2024robustness}.

However, robustness against hardware errors is entirely different from and independent of the robustness of a quantum model against data perturbations, which is the subject of this paper.
The latter type of robustness has been studied in the context of quantum adversarial machine learning~\cite{edwards2020quantum,west2023towards}.
Not surprisingly, just like their classical counterparts, quantum models are also vulnerable to adversarial attacks, both when operating based on classical data~\cite{lu2020quantum,west2023benchmarking} and quantum data~\cite{liu2020vulnerability,lu2020quantum,liao2021robust,du2021quantum,guan2021robustness,weber2021optimal,gong2022universal}.
To mitigate these attacks, it is desirable to design training schemes encouraging adversarial robustness of the resulting quantum model.
Existing approaches in this direction include solving an (adversarial) min-max optimization problem during training~\cite{lu2020quantum} or adding adversarial examples to the training data set~\cite{ren2022experimental}.

Besides robustness, another important aspect of any quantum model is its ability
to generalize to unseen data~\cite{cerezo2022challenges,peters2023generalization}. 
In particular, various works have shown generalization bounds~\cite{abbas2021power,huang2021power,banchi2021generalization,caro2021encoding,jerbi2023shadows}, i.e., bounds on the expected risk of a model depending on its performance on the training data.
While these bounds provide insights into possibilities for constructing quantum models that generalize well, they also face inherent limitations due to their uniform nature~\cite{gil2024understanding}.

\subsection*{Contribution}
This paper presents a flexible and rigorous framework for robustness and generalization of quantum models, providing both a theoretical analysis as well as a simple regularization strategy which allows to systematically adapt robustness and generalization during training (see Figure~\ref{fig:overview} for an overview).
More precisely, we first derive a Lipschitz bound of a given quantum model which explicitly involves the parameters of the data encoding.
Based on this result, we propose a regularized training strategy penalizing the norm of the encoding parameters, which are considered trainable, in order to improve (adversarial) robustness of the model.
Further, we derive a generalization bound which explicitly depends on the parameters of the quantum model and therefore does not share the limitations of existing \emph{uniform} generalization bounds~\cite{gil2024understanding}.
With numerical results, we demonstrate that the proposed Lipschitz bound regularization can indeed lead to substantial improvements in robustness and generalization of quantum models.
Finally, given that the derived Lipschitz bound mainly depends on the norm of the data encoding, our results reveal the importance and benefits of trainable encodings over quantum circuits with a priori fixed encoding as frequently used in variational QML~\cite{schuld2021machine,benedetti2019parameterized,havlicek2019supervised,schuld2019quantum,abbas2021power,schuld2021effect}.

\subsection*{Outline}
The paper is structured as follows.
In Section~\ref{sec:quantum_models}, we introduce the considered class of quantum models with trainable encodings and we state their Lipschitz bound.
Next, in Section~\ref{sec:robustness}, we use the Lipschitz bound to study robustness of quantum models and to derive a regularization strategy for robust training whose benefits are demonstrated with numerical simulations.
We then derive a generalization bound which depends explicitly on the data encoding parameters and we confirm this insight numerically by showing improved generalization under the proposed regularization strategy (Section~\ref{sec:generalization}).
Further, in Section~\ref{sec:trainable}, we discuss an important implication of our results on the benefits of trainable encodings for robustness and generalization.
Finally, Section~\ref{sec:conclusion} concludes the paper.
In the appendix, we provide technical proofs, details on the numerical simulations, as well as additional theoretical and numerical results.

\section{Quantum models and their Lipschitz bounds}\label{sec:quantum_models}

We consider parametrized unitary operators of the form
\begin{align}\label{eq:lipschitz_Ui_def}
    U_{j,\Theta_j}(x)
    =e^{-i(w_j^\top x+\theta_j)H_j},\> j=,1\dots,N
\end{align}
with input data $x \in \mathbb{R}^d$, trainable parameters $\Theta_j=\{w_j,\theta_j\}$, $w_j \in \bbr^d, \theta_j \in \bbr$ and \JB{fixed} Hermitian generators $H_j$.
\JB{The generators $H_j$ are user-chosen, see~\cite{cerezo2022challenges} for references with guidelines.}
Depending on the choice of $H_j$, the operator $U_{j, \Theta_j}$ acts on either one or multiple qubits.
The operators $U_{j,\Theta_j}$ give rise to the parametrized quantum circuit 
\begin{align}\label{eq:lipschitz_U_def}
    U_{\Theta}(x)=U_{N,\Theta_N}(x)\cdots U_{1,\Theta_1}(x),
 \end{align}
where $\Theta=\{\Theta_j\}_{j=1}^N$ comprises the set of trainable parameters.
\JB{Throughout this paper, we abbreviate the $n_{\rmq}$-qubit input state $\ket{0}^{\otimes n_{\rmq}}$ by $\ket0$.}
The quantum model considered in this paper consists of $U_{\Theta}(x)$ applied to $\ket0$ and followed by a measurement w.r.t.\ the observable $\calM$, i.e.,
 \begin{align}\label{eq:QML_circuit}
    f_{\Theta}(x)=\braket{0|U_{\Theta}(x)^\dagger \calM 
    U_{\Theta}(x)|0}.
\end{align}
Note that each of the unitary operators $U_{j,\Theta_j}(x)$ involves the full data vector $x$, i.e., the data are loaded repeatedly into the circuit, a strategy that is commonly referred to as data re-uploading~\cite{perez2020data}.
The encoding of the data $x$ into each $U_{j,\Theta_j}(x)$ is realized via an affine function $w_j^\top x+\theta_j$, where both $w_j$ and $\theta_j$ are trainable parameters.
Hence, we refer to~\eqref{eq:QML_circuit} as a quantum model with trainable encoding.
Such trainable encodings are a generalization of common quantum models~\cite{schuld2021machine,benedetti2019parameterized,havlicek2019supervised,schuld2019quantum,abbas2021power,schuld2021effect}, for which the $w_j$'s are fixed (typically unit vectors) and only the $\theta_j$'s are trained.

Our results rely on Lipschitz bounds~\eqref{eq:def_lipschitz}.
A Lipschitz bound quantifies the maximum perturbation of $f$ that can be caused by input variations.
For the quantum model $f_{\Theta}$, we can state the following Lipschitz bound
\begin{align}\label{eq:thm_lipschitz}
    L_{\Theta}=2\lVert \calM\rVert\sum_{j=1}^N\lVert w_j\rVert\lVert H_j\rVert.
\end{align}
The formal derivation can be found in Appendix~\ref{app:lipschitz_bounds}.
For a given set of parameters $w_j$,~\eqref{eq:thm_lipschitz} allows to compute the Lipschitz bound of the corresponding quantum model.
Note that $L_{\Theta}$ depends only on $w_j$ but it is independent of $\theta_j$.
This fact plays an important role for potential benefits of trainable encodings since the parameters $w_j$ are not optimized during training for fixed-encoding circuits.
We note that all results in this paper hold for arbitrary $p$-norms as long as the same $p$ is used for both vector and induced matrix norms.

\section{Robustness of quantum models}\label{sec:robustness}

Suppose we want to evaluate the quantum model $f_{\Theta}$ at $x$, i.e., we are interested in the value $f_{\Theta}(x)$, but we can only access $f_{\Theta}$ at some \emph{perturbed} input $x'=x+\varepsilon$ with an unknown $\varepsilon$.
Such a setup can arise due to various reasons, e.g., $x$ may be the output of some physical process which can only be accessed via noisy sensors.
The perturbation $\varepsilon$ may also be the result of an adversarial attack, i.e., a perturbation aiming to cause a misclassification by choosing $\varepsilon$ such that 
\begin{align}\label{eq:robustness_adversarial_norm}
    \lVert f_{\Theta}(x+\varepsilon)-f_{\Theta}(x)\rVert
\end{align}
is maximized.
In either case, to correctly classify $x$ despite the perturbation, we require that $f_{\Theta}(x+\varepsilon)$ is close to $f_{\Theta}(x)$, meaning that~\eqref{eq:robustness_adversarial_norm} is small.
According to~\eqref{eq:def_lipschitz}, a Lipschitz bound $L$ of $f_{\Theta}$ quantifies exactly this difference, implying that the maximum possible deviation of $f_{\Theta}(x+\varepsilon)$ from $f_{\Theta}(x)$ is bounded as
\begin{align}\label{eq:analysis_robustness_Lipschitz}
    \lVert f_{\Theta}(x+\varepsilon)-f_{\Theta}(x)\rVert\leq L\lVert \varepsilon\rVert.
\end{align}
This shows that smaller Lipschitz bounds imply better (worst-case) robustness of models against data perturbations.
Thus, using~\eqref{eq:thm_lipschitz}, the robustness of the quantum model $f_{\Theta}$ is mainly influenced by the parameters of the data encoding $w_j$, $H_j$, and by the observable $\calM$.
In particular, smaller values of $\sum_{j=1}^N\lVert w_j\rVert\lVert H_j\rVert$ and $\lVert\calM\rVert$ lead to a more robust model.

We now apply this theoretical insight to train robust quantum models using regularization.
We consider a supervised learning setup with loss $\ell$ and training data set $(x_k,y_k) \in \mathcal{X} \times \mathcal{Y}$ of size $n$.
The following optimization problem can be used to train the quantum model $f_{\Theta}$
\begin{align}\label{eq:training_opt_loss}
    \min_{\Theta} \>\>\frac{1}{n}\sum_{k=1}^{n}\ell(f_{\Theta}(x_k),y_k).
\end{align}
In order to ensure that $f_{\Theta}$ not only admits a small training loss but is also robust and generalizes well, we add a regularization, leading to
\begin{align}\label{eq:training_opt_loss_regularized}
    \min_{\Theta} \>\>\frac{1}{n}\sum_{k=1}^{n}\ell(f_{\Theta}(x_k),y_k)+\lambda\sum_{j=1}^N\lVert w_j\rVert^2\lVert H_j\rVert^2.
\end{align}
Regularizing the parameters $w_j$ encourages small norms of the data encoding and, thereby, small values of the Lipschitz bound $L_{\Theta}$.
\JB{We weight the parameter norms $\lVert w_j\rVert$ by $\lVert H_j\rVert$ due to their joint occurrence in~\eqref{eq:thm_lipschitz}.}
The hyperparameter $\lambda>0$ allows for a trade off between the two objectives of a small training loss and robustness/generalization in the cost function.
Note that the regularization does not involve the $\theta_j$'s since they do not influence the Lipschitz bound~\eqref{eq:thm_lipschitz}, an issue we discuss in more detail in Section~\ref{sec:trainable}.
Moreover, we do not introduce an explicit dependence of the regularization on $\calM$ since we do not optimize over the observable in this paper. We note that penalty terms similar to the proposed regularization can be used for handling hard constraints in binary optimization via the quantum approximate optimization algorithm~\cite{farhi2014quantum,hadfield2017quantum}.
Indeed, the above regularization can be interpreted as a penalty-based relaxation of the corresponding constrained training problem, i.e., of training a quantum model with Lipschitz bound below a specific value.

\begin{figure}[!t]
    \includegraphics[scale=0.50]{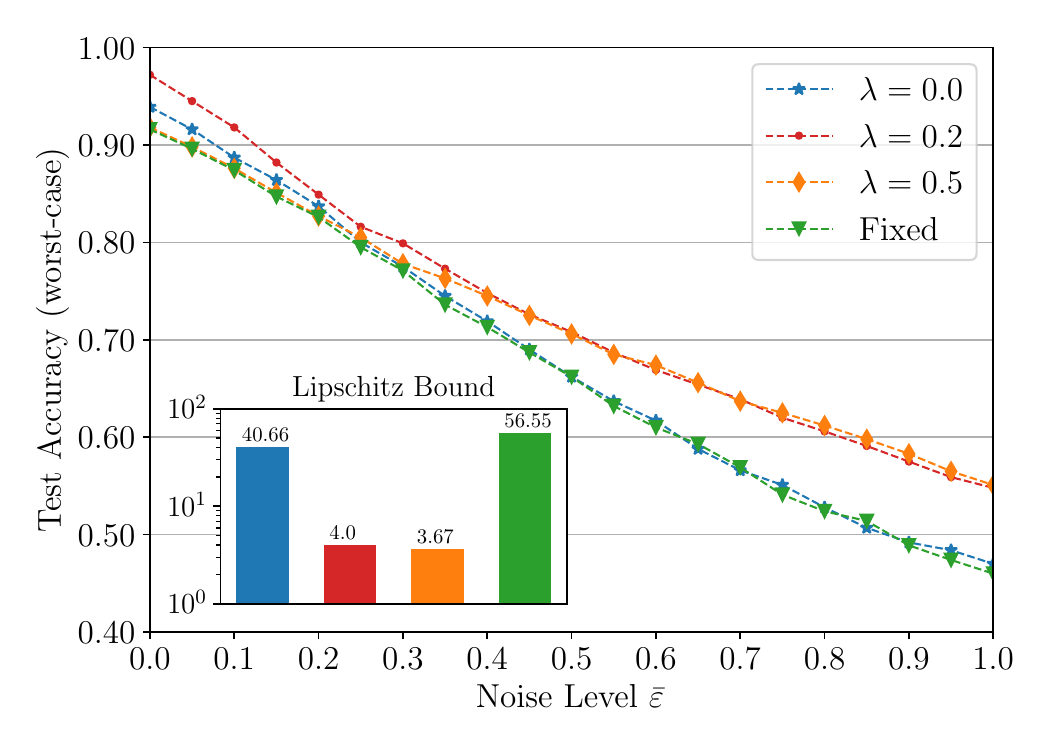}
    \caption{
        We compare robustness of quantum models trained via~\eqref{eq:training_opt_loss_regularized} for $\lambda\in\{0,0.2,0.5\}$ and a quantum model with fixed encoding~\eqref{eq:quantum_model_fixed_encoding}. 
        As training and test set, we draw $n=200$ and $1000$ points $x_i \in \mathcal{X}$, respectively, uniformly at random.
        To study robustness, we perturb each of the $1,000$ test data points by random noise drawn uniformly from $[-\bar{\varepsilon},+\bar{\varepsilon}]^d \ (d=2)$.
        The test accuracy in the plot is the worst case over $200$ noise samples per data point.
        }
    \label{exp:robustness}
\end{figure}

We now evaluate our theoretical findings based on the circle classification problem from~\cite{ahmed2021pennylaneTutorial}:
within a domain of $\mathcal{X} = [-1, +1] \times [-1, +1]$, a circle with radius $\sqrt{2 / \pi}$ is drawn, and all data points inside the circle are labeled with $y = +1$, whereas points outside are labeled with $y = -1$, see Appendix~\ref{app:numerical} (Figure~\ref{exp:data}).
For the quantum model with trainable encoding, we use general SU(2) operators and encode $w_j^\top x + \theta_j$ into the first two rotation angles.
We repeat this encoding for each of the considered $3$ qubits, followed by nearest neighbour entangling gates based on CNOTs.
Such a layer is then repeated $3$ times.
As observable, we use $\mathcal{M} = Z \otimes Z \otimes Z$.
The resulting circuit is illustrated in Appendix~\ref{app:numerical} (Figure~\ref{fig:circuit_3qubits}).
As norms in the regularized training problem~\eqref{eq:training_opt_loss_regularized}, we employ $2$-norms, but we note that exploring different choices is an interesting future direction.
\JB{For example, regularization with non-squared norms (e.g., a $1$-norm) may enforce sparsity of the trained QML model and can, thereby, simplify its implementation on NISQ hardware.}

The numerical results for the robustness simulations are shown in Figure~\ref{exp:robustness}, where we compare the worst-case 
test accuracy 
and Lipschitz bound of three trained models with different regularization parameter $\lambda\in\{0,0.2,0.5\}$.
Additionally, the plot shows the accuracy of a \JB{trained} quantum model with fixed encoding for the same numbers of qubits and layers (see~\eqref{eq:quantum_model_fixed_encoding} and Appendix~\ref{app:numerical} for details).
The worst-case test accuracy of all models is obtained by sampling different noise samples $\varepsilon$ from $[-\bar{\varepsilon},+\bar{\varepsilon}]$.
\JB{
This procedure amounts to finding adversarial noise samples and, therefore, the resulting worst-case test accuracy (approximately) quantifies the adversarial robustness against attacks which are norm-bounded by $\bar{\varepsilon}$.}
As expected, all four models deteriorate with increasing noise level.
For zero noise level $\bar{\varepsilon}=0$, the model with the largest regularization parameter $\lambda=0.5$ (and, hence, the smallest Lipschitz bound $L_{\Theta}=3.67$) has a smaller test accuracy than the non-regularized model with $\lambda=0$.
This can be explained by a decrease in the training accuracy that is caused by the additional regularization in the cost.
For increasing noise levels, however, the enhanced robustness outweighs the loss of training performance and, therefore, the model with $\lambda=0.5$ outperforms the model with $\lambda=0$.
The fixed-encoding model achieves comparable performance to the trainable-encoding model with $\lambda=0.5$ for small noise and the worst performance among all models for high noise.
These observations can be explained by \JB{the high Lipschitz bound of the fixed-encoding model as well as its reduced expressivity, i.e., its limited ability to approximate functions from data due to the fixed encoding parameters $w_j$.}
Finally, the model with $\lambda=0.2$ almost always outperforms the model with $\lambda=0$ and, in particular, it yields a higher test accuracy for small noise levels.
This can be explained by the improved generalization performance caused by the regularization, an effect we discuss in more detail in the following.

\section{Generalization of quantum models}\label{sec:generalization}

The Lipschitz bound~\eqref{eq:thm_lipschitz} not only influences robustness but also has a crucial impact on generalization properties of the quantum model $f_{\Theta}$.
Intuitively, a smaller Lipschitz bound implies a smaller variability of $f_{\Theta}$ and, therefore, reduces the risk of overfitting.
This intuition is made formal via the following generalization bound.

\begin{theorem}\label{thm:generalization_bound}
    (informal version)
    Consider a supervised learning setup with loss $\ell$ and data set $(x_k,y_k)\in\calX\times\calY$ of size $n$ drawn according to the probability distribution $P$.
    For the quantum model $f_{\Theta}$ from~\eqref{eq:QML_circuit}, 
    define the expected risk  $\calR(f_{\Theta})=\int_{\calX\times\calY}\ell(y,f_{\Theta}(x))\rmd P(x,y)$
and the empirical risk $
    \calR_{n}(f_{\Theta})=\frac{1}{n}\sum_{k=1}^{n}\ell(y_k,f_{\Theta}(x_k))$.
    The generalization error of $f_{\Theta}$ is bounded as
    \begin{align}\label{eq:thm_generalization_bound}
        &\Big|\calR(f_{\Theta})-\calR_n(f_{\Theta})\Big|\\\nonumber
        \leq 
        &\,C_1\lVert \calM\rVert\sum_{j=1}^N\lVert w_j\rVert\lVert H_j\rVert+\frac{C_2}{\sqrt{n}}
    \end{align}
    for some $C_1,C_2>0$.
\end{theorem}

The detailed version and proof of Theorem~\ref{thm:generalization_bound} are provided in Appendix~\ref{app:theorem_proof}.
Generalization bounds as in~\eqref{eq:thm_generalization_bound} quantify the ability of $f_{\Theta}$ to generalize beyond the available data.
The bound~\eqref{eq:thm_generalization_bound} depends on the data encoding via $\sum_{j=1}^N\lVert w_j\rVert\lVert H_j\rVert$ and on the observable via $\lVert\calM\rVert$.
In particular, $f_{\Theta}$ achieves a small generalization error if its Lipschitz bound $L_{\Theta}$ is small and the size $n$ of the data set is large.
Note, however, the following fundamental trade-off:
A too small Lipschitz bound $L_{\Theta}$ may limit the expressivity of $f_{\Theta}$ and, therefore, lead to a high empirical risk $\calR_n(f_{\Theta})$, in which case the generalization bound~\eqref{eq:thm_generalization_bound} is meaningless.
In conclusion, Theorem~\ref{thm:generalization_bound} implies a small expected risk $\calR(f_{\Theta})$ if $n$ is large, $L_{\Theta}$ is small, and $f_{\Theta}$ has a small empirical risk $\calR_n(f_{\Theta})$.
In contrast to existing generalization bounds~\cite{abbas2021power,huang2021power,banchi2021generalization,caro2021encoding,jerbi2023shadows}, the bound~\eqref{eq:thm_generalization_bound} is not uniform and explicitly involves the Lipschitz bound~\eqref{eq:thm_lipschitz}, i.e., the parameters of the data encoding.
Hence, Theorem~\ref{thm:generalization_bound} does not share the limitations of uniform QML generalization bounds~\cite{gil2024understanding} \JB{and it can be used to systematically influence the generalization performance during training via regularization.
In particular, according to Theorem~\ref{thm:generalization_bound}, the regularized training problem~\eqref{eq:training_opt_loss_regularized} encourages models with improved generalization properties, where the hyperparameter $\lambda$ trades off between the empirical risk $\calR_{n}(f_{\Theta})$ and the generalization bound~\eqref{eq:thm_generalization_bound}.
Extending our results by studying the direct impact of $\lambda$ on the generalization performance is an interesting next step, which we expect to be non-trivial due to the non-convexity of the loss function in~\eqref{eq:training_opt_loss_regularized}, compare~\cite{huembeli2021characterizing}.
In practice, the hyperparameter $\lambda$ can be tuned, e.g., via cross-validation.
We discuss and interpret the impact of the hyperparameter $\lambda$ in more detail with the following numerical results.
}

\begin{figure}
    \includegraphics[scale=0.50]{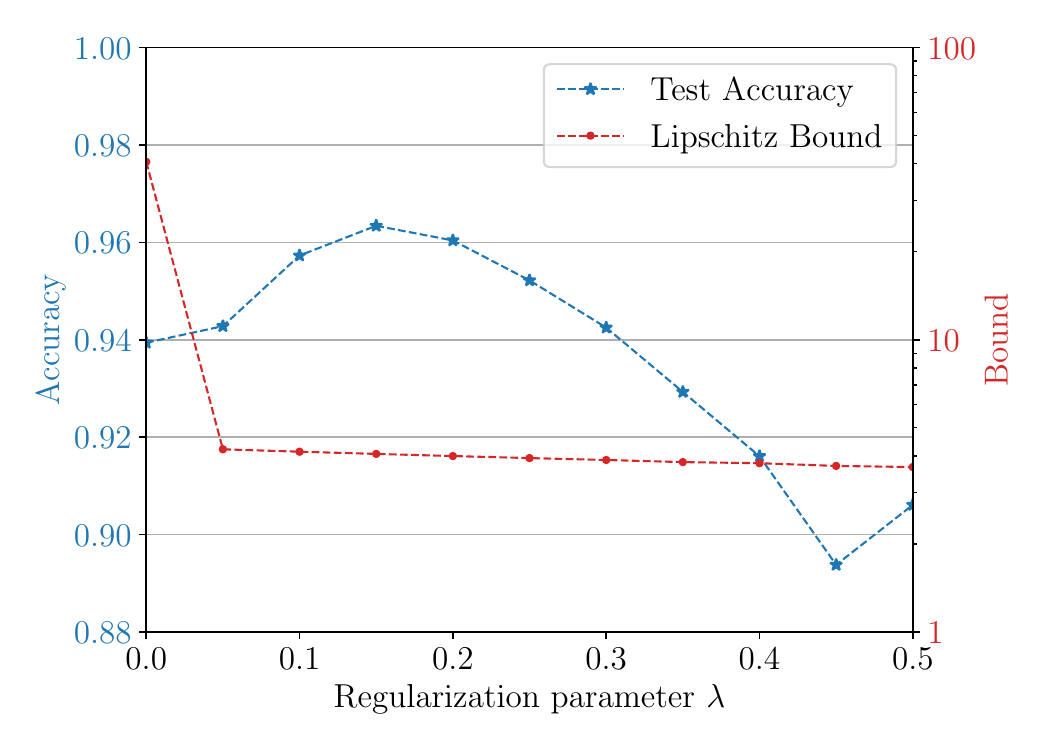}
    \caption{Results for the generalization simulations.
    The training setup is identical to the robustness simulations as described in Figure~\ref{exp:robustness}.
        As test set, we draw $10.000$ points uniformly at random and evaluate the trained models with different regularization parameter $\lambda$.
    }
    \label{exp:generalization}
\end{figure}

We evaluate the generalization performance of the trainable encoding again on the circle classification problem.
The training setup is identical to the robustness simulations and the numerical results are shown in Figure~\ref{exp:generalization}.
Increasing the regularization parameter $\lambda$ decreases the Lipschitz bound $L_{\Theta}$ of the trained model.
In accordance with the generalization bound~\eqref{eq:thm_generalization_bound}, this reduction of $L_{\Theta}$ improves generalization performance with the maximum test accuracy at $\lambda=0.15$.
Beyond this value, the regularization causes a too small Lipschitz bound, limiting expressivity and, therefore, decreasing the training accuracy.
As a result, the test accuracy decreases as well.
This illustrates the role of $\lambda$ as a hyperparameter:
Regularization does not always improve performance, but there is a sweet spot for $\lambda$ at which both superior generalization and robustness over the unregularized setup (i.e., $\lambda=0$) can be obtained.

\section{Benefits of trainable encodings}\label{sec:trainable}

\begin{figure}
    \begin{center}
        \includegraphics[scale=0.7]{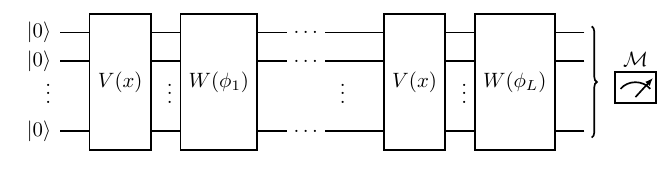}
    \end{center}
    \caption{
        Circuit representation of the quantum model~\eqref{eq:quantum_model_fixed_encoding} with fixed encoding.
        }
    \label{fig:quantum_model_fixed_encoding}
\end{figure}

A popular class of quantum models is obtained by constructing circuits which alternate between data- and parameter-dependent gates, i.e., replacing $U_{\Theta}(x)$ in~\eqref{eq:lipschitz_U_def} by
\begin{align}\label{eq:quantum_circuit_fixed_encoding}
    U^{\mathrm{f}}_{\phi}(x)=W(\phi_L)V(x)\cdots W(\phi_1)V(x),
\end{align}
compare~\cite{schuld2021machine,benedetti2019parameterized,havlicek2019supervised,schuld2019quantum,abbas2021power,schuld2021effect}.
The unitary operators $V, W$ are given by
\begin{align}
    V(x)&=e^{-ix_DG_D}\cdots e^{-ix_{1}G_1},\\
    W(\phi_j)&=e^{-i\phi_{j,p} S_p}\cdots e^{-i\phi_{j,1}S_1}
\end{align}
for trainable parameters $\phi_j$ and generators $G_i=G_i^\dagger$, $S_i=S_i^\dagger$.
The corresponding quantum model is given by
\begin{align}\label{eq:quantum_model_fixed_encoding}
    f^{\mathrm{f}}_{\phi}(x)=\braket{0|U^{\mathrm{f}}_{\phi}(x)^\dagger\calM U^{\mathrm{f}}_{\phi}(x)|0},
\end{align}
see Figure~\ref{fig:quantum_model_fixed_encoding}.

It is not hard to show that the parametrized quantum circuit $U_{\Theta}(x)$ in~\eqref{eq:lipschitz_U_def} generalizes the one in~\eqref{eq:quantum_circuit_fixed_encoding}.
Indeed, $U_{j,\Theta_j}(x)$ in~\eqref{eq:lipschitz_Ui_def} reduces to either
\begin{align}
    e^{-ix_jG_j}\quad\text{or}\quad e^{-i\phi_jS_j}
\end{align}
for suitable choices of $w_j$, $\theta_j$, and $H_j$.
Note that the data encoding of the quantum model $f^{\mathrm{f}}_{\phi}(x)$ is fixed a priori via the choice of $w_j$ and, in particular, it cannot be influenced during training.
Therefore, we refer to $f^{\mathrm{f}}_{\phi}(x)$ as a quantum model with fixed encoding, in contrast to $f_{\Theta}(x)$ in~\eqref{eq:QML_circuit} which contains trainable parameters $w_j$ and, therefore, a trainable encoding.

Benefits of trainable encodings for the expressivity of quantum models have been demonstrated numerically in~\cite{vidal2020input,ovalle2023quantum,jaderberg2024let} and theoretically in~\cite{perez2020data,schuld2021effect}.
In the following, we discuss the importance of trainable encodings for robustness and generalization.
Recall that the Lipschitz bound~\eqref{eq:thm_lipschitz}, which we showed to be a crucial quantifier of robustness and generalization of quantum models, only depends on the observable $\calM$ and on the data encoding $w_j$, $H_j$, but is independent of the parameters $\theta_j$.
Hence, in the quantum model $f_{\phi}^{\mathrm{f}}(x)$ with fixed encoding, the Lipschitz bound~\eqref{eq:thm_lipschitz} cannot be influenced during training and, instead, is fixed a priori via the choice of the Hermitian generators $G_j$.
As a result, training has a limited effect on robustness and generalization properties of fixed-encoding quantum models.

The distinction between trainable and fixed data encodings becomes even more apparent when expressing quantum models as Fourier series~\cite{schuld2021effect}.
In this case, fixed-encoding quantum models choose the frequencies of the Fourier basis functions before the training and only optimize over their coefficients.
On the contrary, trainable-encoding quantum models simultaneously optimize over the frequencies and the coefficients~\cite{jaderberg2024let} which, according to the Lipschitz bound~\eqref{eq:thm_lipschitz}, is key for influencing robustness and generalization properties.

These insights confirm the observation by~\cite{mitarai2018quantum,schuld2020circuit} that fixed-encoding quantum models are neither sensitive to data perturbations nor to overfitting.
On the one hand, resilience against these two phenomena is a desirable property.
However, the above discussion also implies that Lipschitz bound regularization, which is a systematic and effective tool for influencing robustness and generalization~\cite{szegedy2014intriguing,krogh1991simple,hein2017formal,pauli2022training}, cannot be implemented for fixed-encoding quantum models to improve robustness and generalization. 
Indeed, our robustness simulations in Figure~\ref{exp:robustness} show that the fixed-encoding model has a considerably higher Lipschitz bound than all the considered trainable-encoding models.
This implies a significantly worse robustness w.r.t.\ data perturbations and, therefore, leads to a rapidly decreasing test accuracy for larger noise levels.
Further, regularizing the parameters $\phi$ as suggested, e.g., by~\cite{du2021learnability}, does not affect the Lipschitz bound and, therefore, cannot be used to improve the robustness.
In Appendix~\ref{app:numerical}, we study the effect of regularizing the $\phi_j$'s on generalization.
We find that the influence of regularizing the $\phi_j$'s on the test accuracy is limited and likely dependent on the specific ground-truth distribution generating the data and the chosen circuit ansatz.

To conclude, our results show that training the encoding in quantum models not only increases the expressivity but also leads to superior robustness and generalization properties.

\section{Conclusion}\label{sec:conclusion}

In this paper, we studied robustness and generalization properties of quantum models based on Lipschitz bounds.
Lipschitz bounds are a well-established tool in the classical ML literature which not only quantify adversarial robustness but are also closely connected to generalization performance. 
We derived Lipschitz bounds based on the size of the data encoding which we then used to study robustness and generalization of quantum models.
Given that our generalization bound explicitly involves the parameters of the data encoding, it does not face the limitations of uniform generalization bounds~\cite{gil2024understanding}.
Further, our theoretical results highlight the role of trainable encodings combined with regularization techniques for obtaining robust and generalizable quantum models.
The numerical results confirm our theoretical findings, showing the existence of a sweet spot for the regularization parameter for which our training scheme improves both robustness as well as generalization compared to a non-regularized training scheme.
It is important to emphasize that these numerical results with specific choices of rotation and entangling gates are mainly used for illustration, but our theoretical framework applies to all quantum models that can be written as~\eqref{eq:QML_circuit} and, therefore, also allow for different rotation gates, entangling layers, or even parametrized multi-qubit gates.

While our results indicate the potential of using Lipschitz bounds and regularization techniques in QML, it opens up various promising directions for future research.
First and foremost, transferring existing research on Lipschitz bounds in classical ML to the QML setting provides a systematic framework for handling robustness and generalization, beyond the first results presented in this paper.
For example, while we only consider quantum models with affine encodings $w_j^\top x+\theta_j$, it would be interesting to extend our results to more general, nonlinear encodings.
Classical neural networks are ideal candidates for realizing a nonlinear encoding since their Lipschitz properties are well-studied~\cite{szegedy2014intriguing,krogh1991simple,luxburg2004distance,bartlett2017spectrally,neyshabur2017exploring,sokolic2017robust,weng2018evaluating,ruan2018reachability,wei2019data}, which would allow to train hybrid quantum-classical models which are not only expressive but also admit desirable robustness and generalization properties.
Finally, although we focus on variational quantum models, the basic principles of our results are transferrable to different quantum models, including quantum kernel methods~\cite{havlicek2019supervised,schuld2019quantum} or linear quantum models~\cite{jerbi2023quantum}.


\section*{Code Availability}

The source code for the numerical case studies is publicly accessible on GitHub via\\
{\small\url{https://github.com/daniel-fink-de/training-robust-and-generalizable-quantum-models}}.

\section*{Acknowledgment}

This work was funded by Deutsche Forschungsgemeinschaft (DFG, German Research Foundation) under Germany's Excellence Strategy - EXC 2075 - 390740016.
We acknowledge the support by the Stuttgart Center for Simulation Science (SimTech). 
This work was also supported by the German Federal Ministry of Economic Affairs and Climate Action through the project AutoQML (grant no. 01MQ22002A).


\appendix

\section{Lipschitz bounds of quantum models}\label{app:lipschitz_bounds}

In this section, we study Lipschitz bounds of quantum models as in~\eqref{eq:QML_circuit}.
We first derive a Lipschitz bound which is less tight than the one in~\eqref{eq:thm_lipschitz} but can be shown using a simple concatenation argument (Section~\ref{subapp:lipschitz_bounds_simple}).
Next, in Section~\ref{subapp:lipschitz_bounds_proof}, we prove that~\eqref{eq:thm_lipschitz} is indeed a Lipschitz bound.

\subsection{Simple Lipschitz bound based on concatenation}\label{subapp:lipschitz_bounds_simple}

Before stating the result, we introduce the notation
\begin{align}\label{eq:W_Theta_def}
    W=\begin{pmatrix}
        w_1^\top\\\vdots\\w_N^\top
    \end{pmatrix},\>\Omega=\begin{pmatrix}
        \theta_1\\\vdots\\\theta_N
    \end{pmatrix}.
\end{align}

\begin{theorem}
    The following is a Lipschitz bound of $f_{\Theta}$:
    \begin{align}\label{eq:thm_lipschitz_simple}
        L=2\lVert \calM\rVert\lVert W\rVert\sum_{j=1}^N\lVert H_j\rVert.
    \end{align}
\end{theorem}
\begin{proof}
    Our proof relies on the fact that a Lipschitz bound of a concatenated function can be obtained based on the product of the individual Lipschitz bounds.
    To be precise, suppose $f$ can be written as $f=f_1\circ f_2\circ\dots\circ f_h$, where $\circ$ denotes concatenation and each $f_i$ admits a Lipschitz bound $L_i$, $i=1,\dots,h$.
    Then, for arbitrary input arguments $x,y$ of $f$, we obtain
    \begin{align}
        &\lVert f(x)-f(y)\rVert\\\nonumber
        \leq &L_1\lVert f_2\circ\dots\circ f_h(x)-f_2\circ\dots\circ f_h(y)\rVert\\\nonumber
        \leq &\dots\leq L_1L_2\cdots L_h\lVert x-y\rVert.
    \end{align}
    We now prove that~\eqref{eq:thm_lipschitz_simple} is a Lipschitz bound by representing $f_{\Theta}$ as a concatenation of the three functions
\begin{align}\label{eq:thm_lipschitz_proof_g_meas}
    g_{\mathrm{meas}}(z_{\rmm})&=\braket{z_{\rmm}|\calM|z_{\rmm}},
    \\
\label{eq:thm_lipschitz_proof_g_unitary}
g_{\mathrm{unitary}}(z_{\rmu})
&=e^{-iz_{\rmu,N}H_N}\cdots e^{-iz_{\rmu,1}H_1}\ket0,
\\
\label{eq:thm_lipschitz_proof_g_lin}
g_{\mathrm{affine}}(z_{\rma})&=Wz_{\rma}+\Omega.
\end{align}
More precisely, it holds that
    \begin{align}
        f_{\Theta}(x)
        =g_{\mathrm{meas}}\circ 
        g_{\mathrm{unitary}}
        \circ 
        g_{\mathrm{affine}}(x).
    \end{align}
    Hence, any set of Lipschitz bounds $L_{\mathrm{meas}}$, $L_{\mathrm{unitary}}$, 
    $L_{\mathrm{affine}}$ for the three functions $g_{\mathrm{meas}}$, 
    $g_{\mathrm{unitary}}$, $g_{\mathrm{affine}}$ gives rise to a Lipschitz bound of 
    $f_{\Theta}$ as their product:
    \begin{align}\label{eq:thm_lipschitz_proof_L}
        L=L_{\mathrm{meas}}L_{\mathrm{unitary}}L_{\mathrm{affine}}.
    \end{align}
    Therefore, in the following, we will derive individual Lipschitz bounds $L_{\mathrm{meas}}$, $L_{\mathrm{unitary}}$, 
    and $L_{\mathrm{affine}}$.
\\
    \textbf{Lipschitz bound of $g_{\mathrm{meas}}$:}
    Note that 
    \begin{align}
        \frac{\rmd g_{\mathrm{meas}}(z_\rmm)}{\rmd z_\rmm}
        = 2\bra{z_{\rmm}}\calM.
    \end{align}
    Using $\lVert z_{\rmm}\rVert=1$, we infer 
    \begin{align}\label{eq:thm_lipschitz_proof_meas}
        \Big\lVert \frac{\rmd g_{\mathrm{meas}}(z_\rmm)}{\rmd z_\rmm}
        \Big\rVert\leq 2\lVert\calM\rVert.
    \end{align}
    Thus, $L_{\mathrm{meas}}=2\lVert\calM\rVert$ is a Lipschitz bound of $g_{\mathrm{meas}}$.
    \\    
    \textbf{Lipschitz bound of $g_{\mathrm{unitary}}$:}
    It follows from~\cite[Theorem 2.2]{berberich2024robustness} that 
    $L_{\mathrm{unitary}}=\sum_{j=1}^N\lVert H_j\rVert$ is a Lipschitz bound of $g_{\mathrm{unitary}}$.
    \\    
    \textbf{Lipschitz bound of $g_{\mathrm{affine}}$:}
    Given the linear form of $g_{\mathrm{affine}}$, we directly obtain that 
    $L_{\mathrm{affine}}=\lVert W\rVert$ is a Lipschitz bound.
\end{proof}

\subsection{Proof that~\eqref{eq:thm_lipschitz} is a Lipschitz bound}\label{subapp:lipschitz_bounds_proof}

    We first derive a Lipschitz bound on the parametrized unitary $U_{\Theta}(x)$.
    To this end, we compute its differential
    \begin{align}\label{eq:thm_lipschitz_tighter1_proof1} 
        &\rmd U_{\Theta}(x)\\\nonumber 
        =&(\rmd U_{N,\Theta_N}(x))U_{N-1,\Theta_{N-1}}(x)\cdots U_{1,\Theta_1}(x)\\\nonumber
        &+\dots+U_{N,\Theta_N}(x)\cdots U_{2,\Theta_2}(x)(\rmd U_{1,\Theta_1}(x))
    \end{align}
    Note that each term $U_{j,\Theta_j}(x)$ can be written as the concatenation of the two maps
    $g_j$ and $h_j$ defined by
    \begin{align}
        g_j(x)&=w_j^\top x+\theta_j\\
        h_j(z_j)&=e^{-iz_jH_j}.
    \end{align}
    To be precise, it holds that $U_{j,\Theta_j}(x)=h_j\circ g_j(x)$.
    The differentials of the two maps $g_j$ and $h_j$ are given by 
    \begin{align}
        \rmd h_j(z_j)(u)&=-iH_je^{-iz_jH_j}u,\\ \nonumber
        \rmd g_j(x)(v)&=w_j^\top v,
    \end{align}
    where $h_j(z_j)(u)$ denotes the differential of $h_j$ at $z_j$ applied to $u\in\bbr$, and similarly for $g_j(x)(v)$.
    Thus, we have
\begin{align}
    \rmd U_{j,\Theta_j}(x)(v)&=
    (\rmd h_j(g_j(x)))\circ(\rmd g_j(x)(v))\\\nonumber 
    &=-iH_je^{-i(w_j^\top x+\theta_j)H_j}w_j^\top v\\\nonumber 
    &=-iH_j U_{j,\Theta_j}(x)w_j^\top v.
\end{align}
Inserting this into~\eqref{eq:thm_lipschitz_tighter1_proof1}, we obtain 
\begin{align}
    &\rmd U_{\Theta}(x)(v)\\\nonumber 
    =&-i\left( H_NU_{\Theta}(x)w_N^\top v+\dots+U_{\Theta}(x)H_1w_1^\top v\right).
\end{align}
We have thus shown that the Jacobian $J_{\Theta}(x)$ of the map $U_{\Theta}(x)\ket0$ is given by 
\begin{align}
    &J_{\Theta}(x)\\\nonumber 
    =&-i\left( H_NU_{\Theta}(x)\ket0 w_N^\top+\dots+U_{\Theta}(x)H_1\ket0 w_1^\top\right).
\end{align}
Using that $\ket0$ has unit norm, that the $U_{j,\Theta_j}$'s are unitary, as well as the triangle inequality, the norm of $J_{\Theta}(x)$ is bounded as 
\begin{align}
    \lVert J_{\Theta}(x)\rVert\leq \sum_{j=1}^N\lVert w_j\rVert\lVert H_j\rVert.
\end{align}
Thus, $\sum_{j=1}^N\lVert w_j\rVert\lVert H_j\rVert$ is a Lipschitz bound of $U_{\Theta}(x)\ket0$~\cite[p. 356]{apostol1974mathematical}.

Finally, $f_{\Theta}(x)$ is a concatenation of 
$U_{\Theta}(x)\ket0$ and the function $z\mapsto \braket{z|\calM|z}$, which admits the
Lipschitz bound $2\lVert\calM\rVert$, compare~\eqref{eq:thm_lipschitz_proof_meas}.
Hence, a Lipschitz bound of $f_{\Theta}$ can be obtained as the product of these two individual bounds, i.e., as in~\eqref{eq:thm_lipschitz}.$\qquad\qquad\qquad\qquad\qquad\qquad\qquad\qquad\>\>$
$\qed$

\section{Full version and proof of Theorem~\ref{thm:generalization_bound}}\label{app:theorem_proof}
We first state the main result in a general supervised learning setup, before applying it to the quantum model~\eqref{eq:QML_circuit} considered in the paper.
Consider a supervised learning setup with data samples $\{x_k,y_k\}_{k=1}^{n}$ drawn independently and identically distributed from $\calZ\coloneqq\calX\times\calY\subseteq\bbr^d\times\bbr$ according to some probability distribution $P$.
We define the $\epsilon$-covering number of $\calZ$ as follows.\\

\begin{definition}
(adapted from \cite[Definition 1]{xu2012robustness})
    We say that $\hat{\calZ}$ is an \emph{$\epsilon$-cover} of $\calZ$, if, for all $z\in\calZ$, there exists $\hat{z}\in\hat{\calZ}$ such that $\lVert z-\hat{z}\rVert\leq\epsilon$.
    The \emph{$\epsilon$-covering number} of $\calZ$ is
    \begin{align}
        \calN(\epsilon,\calZ)=\min\{|\hat{\calZ}|\mid\hat{\calZ}\>\text{is an $\epsilon$-cover of $\calZ$}\}.
    \end{align}
\end{definition}    

For a generic model $f:\calX\to\calY$, a loss function $\ell:\calY\times\calY\to\bbr$, and the training data $\{x_k,y_k\}_{k=1}^{n}$, we define the expected loss and the empirical loss by 
\begin{align}
    \calR(f)=\int_{\calX\times\calY}\ell(y,f(x))\rmd P(x,y)
\end{align}
and
\begin{align}
    \calR_n(f)=\frac{1}{n}\sum_{k=1}^{n}\ell(y_k,f(x_k)),
\end{align}
respectively.
The following result states a generalization bound of $f$.\\

\begin{lemma}\label{lem:appendix_generalization_bound}
    Suppose
    \begin{enumerate}
        \item[1.] the loss $\ell$ is nonnegative and admits a Lipschitz bound $L_{\ell}>0$,
        \item[2.] $\calZ$ is compact such that the value $M\coloneqq \sup_{y,y'\in\calY}\ell(y,y')$ is finite, and
        \item[3.] $L_f>0$ is a Lipschitz bound of $f$.
    \end{enumerate}
    Then, for any $\gamma,\delta>0$, with probability at least $1-\delta$ the generalization error of $f$ is bounded as
    \begin{align}\label{eq:lem_generalization_bound_suppl}
        &|\calR(f)-\calR_n(f)|\\\nonumber
        \leq &\gamma L_{\ell}\max\{1,L_f\}+M\sqrt{\frac{2\calN(\frac{\gamma}{2},\calZ)\ln2+2\ln(\frac{1}{\delta})}{n}}.
    \end{align}
\end{lemma}
    
\begin{proof}
    For the following proof, we invoke the concept of $(K,\epsilon)$-robustness (adapted from~\cite[Definition 2]{xu2012robustness}):
    The classifier $f$ is \emph{$(K,\epsilon)$-robust} for $K\in\bbn$ and $\epsilon\in\bbr$, if $\calZ$ can be partitioned into $K$ disjoint sets, denoted by $\{C_i\}_{i=1}^K$, such that the following holds:
For all $k=1,\dots,n$, $(x,y)\in\calZ$, $i=1,\dots,K$, if $(x_k,y_k),(x,y)\in C_i$, then 
\begin{align}
    |\ell(y_k,f(x_k))-\ell(y,f(x))|\leq\epsilon.
\end{align}
This property quantifies robustness of $f$ in the following sense:
The set $\calZ$ can be partitioned into a number of subsets such that, if a newly drawn sample $(x,y)$ lies in the same subset as a testing sample $(x_k,y_k)$, then their associated loss values are close.
Let us now proceed by noting that, for any $(x_k,y_k)$, $k=1,\dots,n$, and $(x,y)\in\calZ$, it holds that
    \begin{align}
        &|\ell(y_k,f(x_k))-\ell(y,f(x))|\\\nonumber 
        \leq&L_{\ell}\lVert (y_k,f(x_k))-(y,f(x))\Vert_2\\\nonumber 
        \leq& L_{\ell}\left(\lVert y_k-y\rVert+\lVert f(x_k)-f(x)\rVert\right)\\\nonumber 
        \leq&L_{\ell}\max\{1,L_f\}\lVert (x_k,y_k)-(x,y)\rVert,
    \end{align}
    where we use the Lipschitz bound $L_{\ell}$ of $\ell$, the triangle inequality, and the Lipschitz bound $L_f$ of $f$, respectively.
    Using~\cite[Theorem 6]{xu2012robustness}, we infer that $f$ is $(\calN(\frac{\gamma}{2},\calZ),L_{\ell}\max\{1,L_f\}\gamma)$-robust for all $\gamma>0$.
    It now follows from~\cite[Theorem 1]{xu2012robustness} that, for any $\delta>0$, with probability at least $1-\delta$ inequality~\eqref{eq:lem_generalization_bound_suppl} holds.
\end{proof}

Let us now combine the Lipschitz bound~\eqref{eq:thm_lipschitz} and Lemma~\ref{lem:appendix_generalization_bound} to state a tailored generalization bound for the considered class of quantum models $f_{\Theta}$, thus proving Theorem~\ref{thm:generalization_bound}.

\begin{theorem}\label{thm:generalization_bound_appendix}
    Suppose 
    \begin{enumerate}
        \item[1.] the loss $\ell$ is nonnegative and admits a Lipschitz bound $L_{\ell}>0$ and
        \item[2.] $\calZ$ is compact such that the value $M\coloneqq \sup_{y,y'\in\calY}\ell(y,y')$ is finite.
    \end{enumerate}
    Then, for any $\gamma,\delta>0$, with probability at least $1-\delta$ the generalization error of $f_{\Theta}$ is bounded as
    \begin{align}\label{eq:thm_generalization_bound_suppl}
        &|\calR(f)-\calR_n(f)|\\\nonumber 
        \leq &\gamma L_{\ell}\max\Big\{1,2\lVert \calM\rVert\sum_{j=1}^N\lVert w_j\rVert\lVert H_j\rVert
        \Big\}\\ \nonumber
        &+M\sqrt{\frac{2\calN(\frac{\gamma}{2},\calZ)\ln2+2\ln(\frac{1}{\delta})}{n}}.
    \end{align}
\end{theorem}

Theorem~\ref{thm:generalization_bound_appendix} shows that the size of the data encoding and of the observable directly influences the generalization performance of the quantum model $f_{\Theta}$.
In particular, for smaller values of $\sum_{j=1}^N\lVert w_j\rVert\lVert H_j\rVert$ and $\lVert\calM\rVert$, the expected loss is closer to the empirical loss.
The right-hand side of~\eqref{eq:thm_generalization_bound_suppl} contains two terms:
The first one depends on the parameters of the quantum model $f_{\Theta}$ and characterizes its robustness via the derived Lipschitz bound, whereas the second term decays with increasing data length $n$.
While~\eqref{eq:lem_generalization_bound_suppl} holds for arbitrary values of $\gamma>0$, it is not immediate which value of $\gamma$ leads to the smallest possible bound:
Smaller values of $\gamma$ decrease the first term but increase $\calN(\frac{\gamma}{2},\calZ)$ in the second term (and vice versa).

In contrast to existing QML generalization bounds~\cite{abbas2021power,huang2021power,banchi2021generalization,caro2021encoding,jerbi2023shadows}, Theorem~\ref{thm:generalization_bound_appendix} explicitly highlights the role of the model parameters via the Lipschitz bound.
Using the additional flexibility of the parameter $\gamma$, it can be shown that the generalization bound~\eqref{eq:thm_generalization_bound_suppl} converges to zero when the data length $n$ approaches infinity.
To this end, we use~\cite[Lemma 6.27]{mohri2018foundations} to upper bound the covering number 
\begin{align}\label{eq:suppl_convering_number_upper_bound}
    \calN(\frac{\gamma}{2},\calZ)\leq\left(\frac{6R}{\gamma}\right)^{d+1},
\end{align}
where $R$ is the radius of the smallest ball containing $\calZ$.
Inserting~\eqref{eq:suppl_convering_number_upper_bound} into~\eqref{eq:thm_generalization_bound_suppl} and choosing $\gamma$ depending on $n$ as 
$\gamma=n^{-\frac{1}{2d+2}}$, we infer
\begin{align}
    &|\calR(f)-\calR_n(f)|\\\nonumber
    \leq &\frac{1}{n^{\frac{1}{2d+2}}} L_{\ell}\max\Big\{1,2\lVert \calM\rVert\sum_{j=1}^N\lVert w_j\rVert\lVert H_j\rVert
    \Big\}\\\nonumber
    &+M\sqrt{\frac{2\ln2\left(6R\right)^{d+1}}{\sqrt{n}}+\frac{2\ln(\frac{1}{\delta})}{n}},
\end{align}
which indeed converges to zero for $n\to\infty$.

\section{Numerics: setup and further results}\label{app:numerical}
In the following, we provide details regarding the setup of our numerical results (Section~\ref{suppl_subsec:numerics_setup}) and we present further numerical results regarding parameter regularization in quantum models with fixed encoding (Section~\ref{suppl_subsec:numerics_regularization_fixed_encoding}).

\subsection{Numerical setup}\label{suppl_subsec:numerics_setup}

\begin{figure}
    \begin{center}
    \includegraphics[scale=0.35]{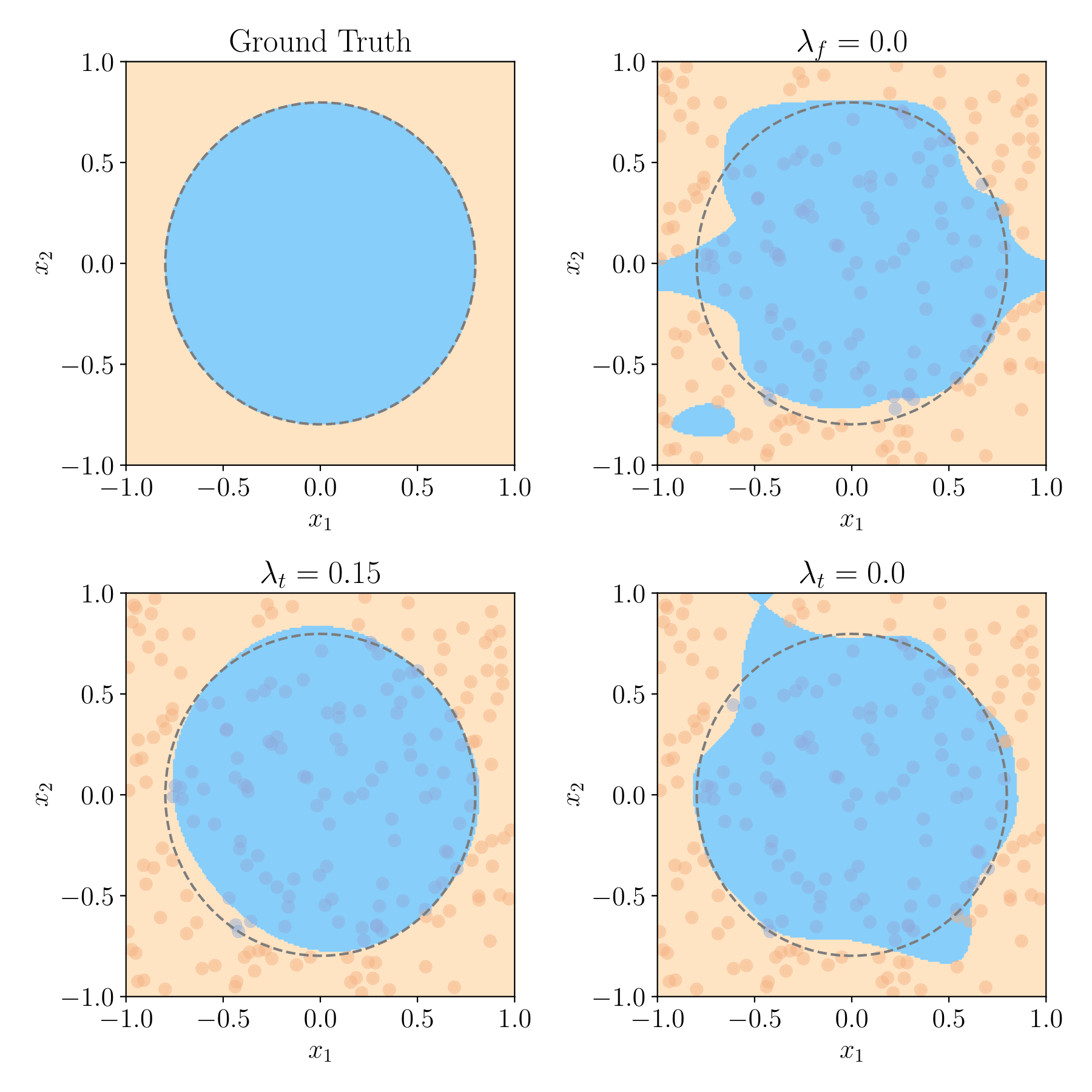}
    \caption{
        From left to right, top to bottom: Illustration of the ground truth of the circle classification problem, the decision boundary for the fixed encoding model with regularization parameter $\lambda_f = 0.0$, the trainable encoding model with $\lambda_t = 0.15$ and with $\lambda_t = 0.0$.
        For the plot, we took the models with the lowest cost over all runs and epochs.
        Furthermore, the small circles denote the $200$ training points.
    }
    \label{exp:data}
    \end{center}
\end{figure}

All numerical simulations within this work were performed using the Python QML library PennyLane~\cite{bergholm2018pennylane}.
As device, we used the noiseless simulator "lightning.qubit" together with the adjoint differentiation method, to enable fast and memory efficient gradient computations.
In order to solve the optimization problem in~\eqref{eq:training_opt_loss_regularized}, we apply the ADAM optimizer using a learning rate of $\eta = 0.1$ and the suggested values for all other hyperparameters~\cite{kingma2014adam}.
Furthermore, we run $200$ epochs throughout and train $12$ models based on different initial parameters for varying regularization parameters $\lambda \geq 0$.
Adding the regularization does not introduce significant computational overhead as the evaluation of the cost only involves a weighted sum of the terms $\| w_j \|^2$.
As final model, we take the set of parameters for the model with minimal cost over all runs and epochs.
Furthermore, the training as well as the robustness and generalization analysis were parallelized using Dask~\cite{dask2016parallelism}.

For the trainable encoding model, the classical data is encoded into the quantum circuit with a general $U_\mathrm{Rot} \in \mathrm{SU(2)}$ unitary parametrized by 3 Euler angles $U_{\mathrm{Rot}}\left(\alpha_j, \beta_j, \gamma_j \right)$ with
\begin{align}\label{sup:rot_angles}
\alpha_j &= w_{j,1}^\top  x + \theta_{j,1} \, , \\
\beta_j &= w_{j,2}^\top  x + \theta_{j,2} \, , \\
\gamma_j &= w_{j,3}^\top  x + \theta_{j,3} \, .
\end{align}
$U_\mathrm{Rot}$ in PennyLane is implemented by the following decomposition
\begin{align}
    &\>U_\mathrm{Rot}(\alpha, \beta, \gamma) = R_Z(\gamma) R_Y(\beta) R_Z(\alpha) \, , \\\nonumber
&= \begin{pmatrix}
e^{-\frac{i}{2}(\gamma + \alpha)} \cos(\frac{\beta}{2}) & -e^{\frac{i}{2}(\gamma - \alpha)} \sin(\frac{\beta}{2}) \\
e^{-\frac{i}{2}(\gamma - \alpha)} \sin(\frac{\beta}{2}) & e^{\frac{i}{2}(\gamma + \alpha)} \cos(\frac{\beta}{2})
\end{pmatrix} \, .
\end{align}

\begin{figure}
    \begin{center}
    \includegraphics[scale=0.50]{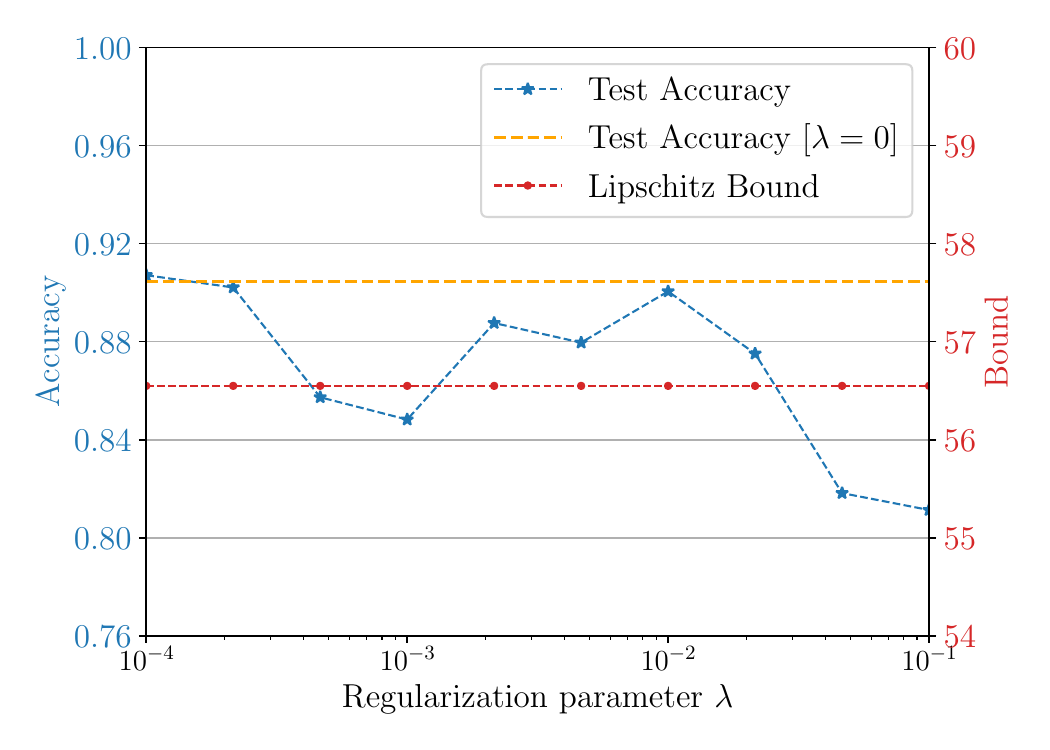}
    \caption{Results for the generalization simulations for the circle classification problem when using the fixed-encoding quantum model~\eqref{eq:quantum_model_fixed_encoding} and regularization of the parameters $\phi$.
        The training and test setup are identical to simulations shown in Figures~\ref{exp:robustness} and~\ref{exp:generalization}.
        We plot the dependency of the test accuracy and the Lipschitz bound on the hyperparameter $\lambda$ entering the regularized training problem~\eqref{eq:training_opt_loss_regularized_fixed}.
        Further, we plot the test accuracy for the best solution of the unregularized training problem.
    }
    \label{exp:generalization_fixed_encoding}
    \end{center}
\end{figure}

\begin{figure*}
	\begin{center}
	\includegraphics[scale=0.73]{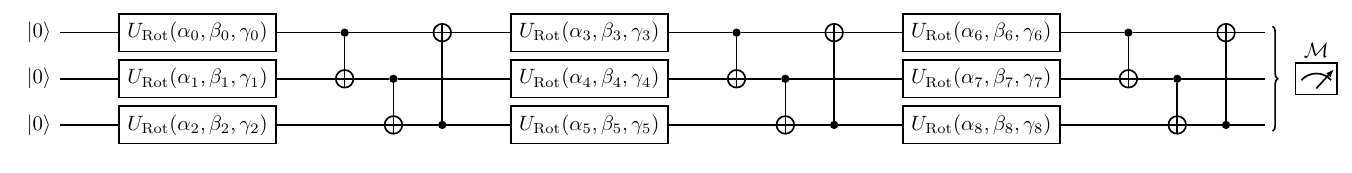}
	\caption{
        The trainable encoding quantum model employed in our numerical case studies uses general $U_\mathrm{Rot} \in \mathrm{SU}(2)$ unitaries parametrized by three Euler angles $\alpha_i$, $\beta_i$, $\gamma_i$, that each have the form $w_i^\top x+\theta_i$ for trainable parameters $w_i$, $\theta_i$.
        We set $\gamma_j = 0$, $j=0,\dots,8$, to enable a fair comparison to a quantum model with fixed encoding (since the data are two-dimensional, only two angles are needed for encoding the data via the latter). 
    }
	\label{fig:circuit_3qubits}
	\end{center}
\end{figure*}

In our numerical case study we set $\gamma_j = 0$ for all $j$ since this 1) still allows to reach arbitrary points on the Bloch sphere and 2) enables an easier comparison to fixed-encoding quantum models.
In order to introduce entanglement in a hardware-efficient way, we use a ring of CNOTs.
The considered circuit is shown in Figure~\ref{fig:circuit_3qubits} and involves three layers of rotations and entanglement, which we observed to be a good trade-off between expressivity and generalization.
More precisely, in our simulations, fewer layers were not sufficient to accurately solve the classification task, whereas more layers led to higher degrees of overfitting. 
For the fixed-encoding quantum models as in~\eqref{eq:quantum_model_fixed_encoding}, we use a similar $3$-layer ansatz, encoding the two entries of the 2D data points into the first and second angle $\alpha_j$, $\beta_j$ of the rotation gates, followed by a parametrized  $\mathrm{SU(2)}$ rotation with three free parameters \JB{that are optimized, compare $\phi_j$ in~\eqref{eq:quantum_circuit_fixed_encoding}.}
We also perform a classical data pre processing, scaling the input domain from $[-1, +1]$ to $[-\pi, +\pi]$, such that the full possible range of the rotation angles can be utilized.

In Figure~\ref{exp:data}, we plot the ground truth and the decision boundaries for the two quantum models corresponding to $\lambda = 0.0$ and $\lambda = 0.15$, as well as the decision boundary for the fixed-encoding model.
As expected, the decision boundary resulting from the regularized training is significantly smoother than the unregularized one, explaining the superior robustness and generalization of the former.
Further, the fixed-encoding model does not accurately capture the ground truth due to its limited expressivity and high Lipschitz bound.

\subsection{Regularization in quantum models with fixed encoding}\label{suppl_subsec:numerics_regularization_fixed_encoding}

In the main text, we have seen that the Lipschitz bound of the quantum model with fixed encoding $f_{\phi}^\rmf(x)$ in~\eqref{eq:quantum_model_fixed_encoding} cannot be adapted by changing the parameters $\phi$.
As a result, it is not possible to use Lipschitz bound regularization for improving robustness and generalization.
In the following, we investigate whether regularization of $\phi$ can instead be used to improve generalization performance. 
More precisely, we consider the same numerical setup as for our generalization results with trainable encoding depicted in Figure~\ref{exp:generalization}.
The main difference is that we consider a fixed-encoding quantum model $f_{\phi}^\rmf(x)$ (compare Figure~\ref{fig:quantum_model_fixed_encoding}) which is trained via the following regularized training problem 
\begin{align}\label{eq:training_opt_loss_regularized_fixed}
    \min_{\phi} \>\>\frac{1}{n}\sum_{k=1}^{n}\ell(f_{\phi}^\rmf(x_k),y_k)+\lambda\sum_{j=1}^L\lVert \phi_j\rVert^2.
\end{align}

The regularization with hyperparameter $\lambda>0$ aims at keeping the norms of the angles $\phi_j$ small.
Figure~\ref{exp:generalization_fixed_encoding} depicts the test accuracy and Lipschitz bound of the resulting quantum model for different regularization parameters $\lambda$.
First, note that the Lipschitz bound is indeed constant for all choices of $\lambda$ due to the fixed encoding.
Comparing Figures~\ref{exp:generalization} and~\ref{exp:generalization_fixed_encoding}, we see that the trainable encoding yields a significantly higher test accuracy in comparison to the fixed encoding.
Moreover, the influence of the regularization parameter on the test accuracy is much less pronounced for the fixed encoding than for the trainable encoding, confirming our previous discussion on the benefits of trainable encodings.
The test accuracy is not entirely independent of $\lambda$ since 1) regularization of the parameters $\phi_j$ influences the optimization and can improve or deteriorate convergence and 2) biasing $\phi$ towards zero may be beneficial if the underlying ground-truth distribution is better approximated by a quantum model with small values $\phi$.
Whether 2) brings practical benefits is, however, highly problem-specific as it depends on the distribution generating the data and on the circuit ansatz.

\bibliography{Literature}

\end{document}

%% file: commands_environments.tex
\newtheorem{theorem}{\bf Theorem}[section] \newtheorem{definition}{\bf Definition}[section]
\newtheorem{lemma}{\bf Lemma}[section]

\usepackage{bbm}

\newcommand\rma{\mathrm{a}}

\newcommand\rmd{\mathrm{d}}

\newcommand\rmf{\mathrm{f}}

\newcommand\rmm{\mathrm{m}}

\newcommand\rmq{\mathrm{q}}

\newcommand\rmu{\mathrm{u}}

\newcommand\bbn{\mathbb{N}}

\newcommand\bbr{\mathbb{R}}

\newcommand\calM{\mathcal{M}}
\newcommand\calN{\mathcal{N}}

\newcommand\calR{\mathcal{R}}

\newcommand\calX{\mathcal{X}}
\newcommand\calY{\mathcal{Y}}
\newcommand\calZ{\mathcal{Z}}
